\documentclass[twocolumn,prb,amsmath,amssymb,verbatim,showpacs]{revtex4}
\usepackage[latin9]{inputenc}
\usepackage{textcomp}
\usepackage{mathrsfs}
\usepackage{amsmath}
\usepackage{amssymb}
\usepackage{graphicx}
\usepackage{txfonts}

\begin{document}
\title{Thermally isolated Luttinger liquids with noisy Hamiltonians}
\author{Luca D'Alessio}
\affiliation{Physics Department, Boston University, Boston, Massachusetts 02215, USA}
\author{Armin Rahmani}
\affiliation{Theoretical Division, T-4 and CNLS, Los Alamos National Laboratory,
Los Alamos, New Mexico 87545, USA}

\date{\today}
\begin{abstract}
We study the dynamics of a quantum-coherent thermally isolated Luttinger
liquid with noisy Luttinger parameter. To characterize the fluctuations of the absorbed energy in generic noise-driven systems,  we first identify
two types of energy moments, which can help tease apart the effects
of classical (sample-to-sample) and quantum sources of fluctuations. One type of moment captures the total fluctuations due to both sources, while the other one captures the effect of the classical source only. We then demonstrate that in the Luttinger liquid case, the two types of moments agree in the thermodynamic limit, indicating that the classical source dominates. In contrast to equilibrium thermodynamics,  in this driven system the relative fluctuations of energy do not decay with the system size. Additionally, we study the deviations of equal-time correlation functions from their ground-state value, and find a simple scaling behavior.

\end{abstract}
\maketitle

\section{Introduction\label{sec:intro}}

Recent experimental developments with ultracold atoms have motivated
numerous studies of the nonequilibrium dynamics of thermally isolated
many-body quantum systems. Most of these studies focus on \textit{deterministic}
quantum evolution, generated, e.g., by a sudden quench or gradual
ramping of the Hamiltonian (see Ref.~\onlinecite{Polkovnikov_Rev} and
the references therein). However, \textit{stochastic} driving of thermally
isolated systems with noisy Hamiltonians~\cite{Bunin11,Silva,Pichler2012}
is much less studied, and, in particular, the role of quantum coherence
remains largely unexplored.

Understanding the unitary dynamics generated by stochastic Hamiltonians
is of interest from both experimental and fundamental viewpoints.
On the experimental side, the preparation of strongly correlated ground-state wave functions by controlled unitary evolution
(for the purpose of quantum simulations, for example) is an important goal in cold-atom physics. However, in any real experiment, the implementation
of the desired (time-independent) Hamiltonian is not perfect and noisy fluctuations
are unavoidable. Additionally, prescribed time-dependent protocols (such as, e.g., complex optimal control protocols~\cite{Chen10,Doria11,Rahmani11,Rahmani12}) have inaccuracies, which may be in the form of noisy fluctuations.
In such cases, the stochasticity has a detrimental effect, which one
needs to minimize. An in-depth understanding of the effect of noisy
Hamiltonian evolution is thus crucial for, on one hand, correctly
predicting the results of experiments and, on the other hand, designing
sophisticated experimental setups, which are robust against the
effect of the noise. \cite{Pichler2012}

On the fundamental side, evolution with noisy Hamiltonian is a natural extension of the physics of disorder to the time domain. Disorder (in real space) has been widely studied for
time-independent Hamiltonians starting from the pioneering work of
Anderson, \cite{Anderson} and has turned into a rich area of research. The advances in nonequilibrium dynamics motivate the study of the effect of disorder in time for time-dependent Hamiltonians.
To address this question, it is necessary to analyze an ensemble of
unitary evolutions characterized by different realization of noise
and develop methods for computing ensemble-averaged quantities.

Another new question of fundamental interest is understanding the role of different sources of fluctuations.  In stochastically driven systems, the fluctuations of physical observables
stem from two distinct sources: (i) the classical (sample-to-sample) stochastic nature
of the driving (different realizations of noise result in different
wave functions) and (ii) inherent quantum fluctuations (each wave function
can be a coherent superposition of eigenstates). Our objective in
this paper is to (i) characterize the effect of the two sources above
on energy fluctuations, and (ii) understand the effects of such stochastic
driving on correlation functions.

In this paper, we consider the following general setup: a quantum
system with a local Hamiltonian $H(g)$, which depends on some parameter
$g$ (such as for example a coupling constant). We assume the system
is initially in the ground state of $H_{0}\equiv H(g_{0})$. For $t>0$,
the parameter $g$ fluctuates in time: $g(t)=g_{0}+\delta g(t)$,
and, by assumption, $|\delta g(t)|\ll|g_{0}|$. For each realization
$\delta g(t)$ of noise, the system is then described by a pure-state
wave function, which evolves unitarily with Hamiltonian $H(g(t))$ (generically $H(g(t))$ does not commute with $H_0$).
Thus, the quantum
evolution creates excitations with respect to the ground state of
$H_{0}$, increasing the system's energy. The driving is assumed external,
i.e. there is no feedback action from the quantum system on $\delta g(t)$.
Therefore, in isolation from a thermal environment (without a mechanism
for dissipation), the system can absorb energy \textit{ad infinitum}.
Such noisy systems can, for example, be realized by cold atoms in
optical potentials, with $\delta g(t)$ generated by small fluctuations
in the optical potential. Note that this setup can be readily generalized
to the case of several fluctuating parameters. 
\begin{figure}
\includegraphics[width=0.9\columnwidth]{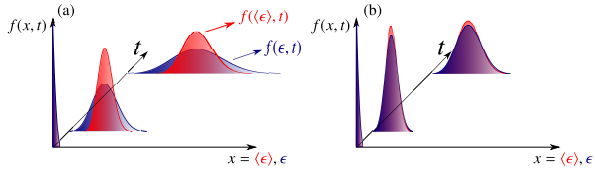} \caption{(Color online) Distribution functions (over noise) of the measured
energy $\epsilon$ and its quantum expectation value $\langle\epsilon\rangle$.
(a) Generically, the former is wider than the latter (the difference
between the two serves as a diagnostic for the significance of quantum
fluctuations). (b) For a Luttinger liquid in the thermodynamic limit,
the two distribution functions are close (classical fluctuations dominate).}
\label{fig1} 
\end{figure}

The absorbed energy of such noise-driven thermally isolated systems
is a \textit{time-dependent} random variable $\epsilon$ (as opposed
to the steady states which emerge in noise-driven systems coupled
to a heat bath.~\cite{Dallatorre}) What is the precise meaning of
$\epsilon$ for systems described by coherent superpositions of energy
eigenstates? How can we characterize the average ${\rm E}(\epsilon)$
and the variance ${\rm Var}(\epsilon)$ of this random variable as
a function of time? How are the correlation functions affected by
this stochastic driving?

To answer the above questions, we first identify two types of quantum-
and noise-averaged moments of energy, which make the notion of a random
absorbed energy precise, and, help tease apart the effects of the
classical and quantum sources of fluctuations discussed above (see
Fig.~\ref{fig1}). We then perform explicit analytical calculations
of these moments, as well as different noise-averaged correlation
functions, for generic one-dimensional systems described by the Luttinger
liquid (LL) theory. The nonequilibrium dynamics of Luttinger liquids (due to deterministic protocols such as interaction quenches) has been a subject of intense studies with various methods.~\cite{Calabrese2006,Cazalilla2006,Manmana2007,Calabrese2007,Calabrese2009,Rigol2009,Mitra2011,Dora2011,Mitra2012,Meden2012,NJP,Dora2013, Dora2013b} In this work, we consider a LL Hamiltonian with a fluctuating LL parameter, i.e., $H(K(t))$, and use a method based on the (first-quantized) evolution of momentum-mode wave functions, which proves extremely powerful in the analysis of stochastic driving. 

Our main results are as follows. We find an exact analytical expression for the noise- and quantum-averaged absorbed energy:
\begin{equation}
\overline{\langle\epsilon\rangle}=\frac{L}{8\pi K_{0}^{2}W^{2}t}\left(e^{2K_{0}^{2}\pi^{2}W^{2}t}-2\pi^{2}K_{0}^{2}W^{2}t-1\right),
\end{equation}
where $L$ is the system size, $t$ is the time since the beginning of the evolution, and $W^{2}$
is a constant with dimension of time which characterizes the strength
of noise in the Luttinger parameter: $1/K(t)=1/K_0+\delta \alpha(t)$ 
with $\overline{\delta \alpha(t_1)\delta \alpha(t_2)}=W^2 \delta(t_1-t_2)$.
Here $1/K_0$ and $\delta \alpha(t)$ correspond respectively to $g_0$ and $\delta g(t)$ introduced earlier.
We find that, in the thermodynamic
limit, the dominant contribution to the energy fluctuations stems from the classical source. In
the thermodynamic limit, and in the regime of validity of the LL description, we
find a general relationship (independent of the strength of noise
and the Luttinger parameter) between the average and the variance
of energy: 
\begin{eqnarray}
{\rm Var}(\epsilon)={\cal F}(\pi t)\left[{\rm E}(\epsilon)\right]^{2},\label{eq:main1}
\end{eqnarray}
 where ${\cal F}(x)$ is a dimensionless function of $x=\pi tu/a$
(with velocity $u$ and lattice spacing $a$ set to unity), which
decays as $1/x$ for large $x$. In contrast to equilibrium thermodynamics, where the relative fluctuations$\sqrt{{\rm Var}(\epsilon)}/{\rm E}(\epsilon)$ scale as $1/\sqrt{L}$, in this case, the relative fluctuations are independent of the systems size and, instead, die off as $1/\sqrt{t}$ for large $t$. 

We also evaluate certain noise-averaged
equal-time correlation functions, and find that in the limit of small absorbed energies, correlation functions scaling as $x^{-\Delta}$ in the ground state, deviate from their ground-state value by amounts proportional to $W^2 t/x^{\Delta+2}$. Interestingly, we are able to find
an exact analytical result for the current-current correlation function,
\begin{equation}
\overline{{\cal C}(x)}=\frac{K_{0}}{2\pi}\int_{0}^{\pi}dq\, q\,\cos\left(q\, x\right)\exp\left(2K_{0}^{2}W^{2}q^{2}t\right),
\end{equation}
which can be written in closed form in terms of the error functions.

\section{Two types of energy moments \label{sec:moments} }

Let us now discuss the two aforementioned moments of energy. As the
evolution with different noise realizations can lead to different
pure-state wave functions $\psi$ at time $t$, it is helpful to introduce a wave-function
probability distribution $f(\psi,t)$, which encodes this stochastic effect. \cite{note}
The inherent quantum fluctuations, on the other hand, stem from the
internal structure of the wave functions, which are, generically,
coherent superpositions of all energy eigenstates: $|\psi\rangle=\sum_{n}c_{n}^{\psi}|n\rangle$,
where $H_{0}|n\rangle=\epsilon_{n}|n\rangle$. At $t=0$, the system
is in the ground state of $H_{0}$ so $f(\psi,0)$ is a delta function
(there are no classical fluctuations). Additionally, since the ground
state is an eigenstate of the energy, there are no quantum fluctuations
either at $t=0$. For $t>0$ the evolution both broadens the distribution
$f(\psi,t)$ (classical source), and makes the wave function $|\psi\rangle$
a coherent superposition of multiple eigenstates (quantum source), giving rise
to (total) energy fluctuations which originate from both sources above.

Experimentally, the outcomes of an energy measurement at
time $t$ (with respect $H_{0}$ \cite{H0}) is described by the distribution
function $f(\epsilon,t)$, which mixes both classical and quantum
contributions (see Fig.~\ref{fig1}). We can characterize $f(\epsilon,t)$
through the moments $\overline{\langle\epsilon^{m}\rangle}=\int d\epsilon f(\epsilon)\epsilon^{m}$,
where the brackets (overline) indicate a quantum (noise) average.
Moreover, these moments can be written as $\overline{\langle\epsilon^{m}\rangle}=\sum_{n}\epsilon_{n}^{m}{\cal P}(\epsilon_{n})$,
where ${\cal P}(\epsilon_{n})=\int d\psi f(\psi,t)|c_{n}^{\psi}|^{2}$
is the probability of measuring $\epsilon_{n}$, which yields 
\begin{equation}
\overline{\langle\epsilon^{m}\rangle}=\int d\psi f(\psi,t)\langle\psi|H_{0}^{m}|\psi\rangle={\rm tr}\left[H_{0}^{m}\rho(t)\right],\label{average2}
\end{equation}
 where $\rho(t)\equiv\int d\psi f(\psi,t)|\psi\rangle\langle\psi|$
is the density matrix at time $t$.

To separate the contributions of the two sources above, we can consider
the fluctuations of the expectation value of energy, $\langle\epsilon\rangle\equiv\langle\psi|H_{0}|\psi\rangle$,
over different realizations of noise, which are characterized by the
following moments: 
\begin{equation}
\overline{\langle\epsilon\rangle^{m}}=\int d\langle\epsilon\rangle f(\langle\epsilon\rangle,t)\langle\epsilon\rangle^{m}=\int d\psi f(\psi,t)\left(\langle\psi|H_{0}|\psi\rangle\right)^{m}.\label{average1}
\end{equation}
 As $\langle\epsilon\rangle$ is a quantum-averaged quantity, its
fluctuations stem solely from the classical stochastic driving. Thus,
the difference between these two types of moments {[}Eqs. \eqref{average2}
and \eqref{average1}{]} can serve as a theoretical diagnostic for
the relative importance of the two sources of fluctuations (see Fig.~\ref{fig1}).
Note that the moments~\eqref{average1} can not be written in terms
of the density matrix for $m>1$.~\cite{Molmer,Petruccione} Fluctuations
of quantities other than energy can be similarly studied by expanding
the wave function in the corresponding eigenbasis.

\section{Wave-function method}

Let us now turn to the specific model studied in this paper, namely,
a Luttinger liquid, which is a universal low-energy description of
interacting fermions and bosons in one dimension. \cite{LL-universal-1}
Since a noise-driven dissipationless system can keep absorbing energy,
this low-energy description will eventually break down for most experimentally
realistic scenarios. In this paper, we focus on dynamics over a finite
time scale where the LL description remains valid. Notice that such
time scales can be extended by decreasing the strength of noise. There are numerous proposals for realizing the LL physics (with negligible coupling to the environment) with both bosonic and fermionic atoms. \cite{Olshanii,Petrov,LL_th_exp_1,LL_th_exp_2} Luttinger liquids have already been realized with bosonic atoms in the Tonks-Girardeau gas,~\cite{LL_exp1,LL_exp2} and in elongated quasicondensates.~\cite{Gorlitz,Dettmer,Richard,Jo,Hofferberth,Schmiedmayer,kitagawa-science}

In terms of the Luttinger parameter $K$ and velocity $u$, the LL
Hamiltonian is given by 
\begin{equation}
H(K)=u\sum_{q>0}\left(K\:\Pi_{q}\Pi_{-q}+\frac{1}{K}\: q^{2}\:\Phi_{q}\Phi_{-q}\right),\label{eq:H-momentum-space}
\end{equation}
 where $\Phi_{q}$ are bosonic fields and $\Pi_{q}$ their conjugate
momenta. The Hamiltonian above can be written as $\sum_{q>0}\left(H_{q}^{\Re}+H_{q}^{\Im}\right)$,
where $H_{q}^{\Re\:(\Im)}$ is the Hamiltonian of a single harmonic
oscillator involving only the real (imaginary) part of $\Phi_{q}$.
It is convenient to shift the Hamiltonian by a constant, i.e., $H_{q}^{\Re\:(\Im)}\rightarrow H_{q}^{\Re\:(\Im)}-uq/2$,
so that the energies are measured with respect to the ground state.
We consider a system initially in the ground state of $H_{0}=H(K_{0})$,
which evolves with $H(K(t))=H(K_{0}+\delta K(t))$ for $t>0$, where
$\delta K(t)\ll K_{0}$ represents the noise. As the fluctuations
of velocity $u$ correspond to a trivial rescaling of the Hamiltonian,
we set $u$ to unity throughout this paper.

Expanding the Hamiltonian~\eqref{eq:H-momentum-space} in $\delta K$
results in quadratic (in the bosonic fields) noise terms. Thus, integrating
out the noise at the outset (as in Refs.~\onlinecite{Silva,Dallatorre,Wilson2012})
leads to an interacting (quartic) effective action, which is difficult
to treat exactly. In our case it is convenient to use an alternative
approach, i.e., the wave-function approach, which consists of the following
steps. (i) We parametrize the many-body time-dependent wave function
of the system with complex numbers $z_{q}$. (ii) We transform the
time-dependent Schr\"odinger equation to an equation of motion for the
parameters above, and then expand these equations in the noise terms
to obtain a stochastic Langevin equation. (iii) We express the observables
of interest in terms of the parameters above, and study their stochastic
dynamics with the Langevin equation. This method allows us to reduce
the quantum dynamics in Hilbert space to a set of equations of motion
(in our case Langevin equations) for the parameters. Generically, the
number of such parameters grows exponentially with system size, but
for exactly solvable models such as Luttinger liquids a much smaller number of parameters may be
necessary.

For the Luttinger liquid above, the Luttinger parameter $K(t)$ is
assumed spatially uniform. Therefore, momentum is a good quantum number
throughout the noisy evolution. Moreover, the ground state of $H_{0}$
is a direct product of Gaussian wave functions for different momentum
modes $q$. Since each mode evolves with a quadratic Hamiltonian,
the time-dependent wave function of the system, i.e., the solution
of the time-dependent Schr\"odinger equation, retains the form 
\begin{equation}
\Psi(\{\Phi_{q}\},t)=\prod_{q>0}\left(\frac{2\: q\left[\Re\: z_{q}(t)\right]}{\pi}\right)^{\frac{1}{2}}\exp\left[-q\: z_{q}(t)\:|\Phi_{q}|^{2}\right],\label{eq:psi}
\end{equation}
where the parameter $z_{q}(t)$ now satisfies the Riccati equation
$i\dot{z}_{q}(t)=\frac{q}{K(t)}\left\{[K(t)\, z_{q}(t)]^{2}-1\right\}$,
with initial condition $z_{q}(0)=K_{0}^{-1}$.~\cite{Polkovnikov08,Rahmani11}
Note that the number of these parameters goes as the number of modes
$q$, which in turn scales linearly with the system size. We also
mention in passing that the single-mode wave functions $\Psi(\Phi_{q},t)$
are exponentially localized on low-energy levels (see Appendix\ref{sec:Eigenstate-expansion-of}),
thus generating an effective high-energy cutoff. 

By expanding the above Riccati equation in $\delta K$, we then obtain
the following nonlinear Langevin equation: 
\begin{equation}
i\dot{z}_{q}=\frac{q}{K_{0}}\left(K_{0}^{2}z_{q}^{2}-1\right)-q\left(K_{0}^{2}z_{q}^{2}+1\right)\delta\alpha,\label{eq:lang}
\end{equation}
where $\delta\alpha(t)=-\delta K(t)/K_{0}^{2}$. Note that there is
no dissipative term in the above equation. Also notice that although
different modes are decoupled in the Langevin equations above, different
$z_{q}(t)$ evolve with the \textit{same} $\delta\alpha(t)$. Therefore
the many-mode quantities must be computed by taking the noise-induced
correlations into account.

We now assume Gaussian noise with zero average and second moments
characterized by strength $W$ and correlation time $\tau$ as in
the Ornstein-Uhlenbeck process: 
\begin{equation}
\overline{\delta\alpha(t_{1})\delta\alpha(t_{2})}=\frac{W^{2}}{2\tau}e^{-|t_{1}-t_{2}|/\tau}.\label{eq:corr_noise}
\end{equation}
In the limit $\tau\rightarrow0^{+}$, the right-hand side of Eq.~(\ref{eq:corr_noise})
reduced to $W^{2}\delta(t_{1}-t_{2})$ describing Gaussian white noise.
We leave a discussion of the effects of colored noise (finite $\tau$)
to Appendix \ref{sec:Effects-of-colored}, and only consider the $\tau\rightarrow0^{+}$
limit here. As the white noise is thought of as a limit of a continuous
process, we use the Stratonovich interpretation for Eq.~(\ref{eq:lang}).
\cite{Koeden} The stochastic differential equation described by
Eqs.~(\ref{eq:lang}) and (\ref{eq:corr_noise}) is one of the key
equations of this paper: it governs the stochastic evolution of $z_{q}(t)$,
which in turn determines the many-body wave function, and, consequently,
all the observables of the system. 

Several observables can be simply written in terms of $z_{q}$. For instance,
the first moment of the absorbed energy and the equal-time correlation function
of the bosonic fields are respectively given by (see Appendix\ref{sec:Expression-for-observables}):
\begin{eqnarray}
&&\langle H_{q}(K_{0})\rangle=\frac{q}{2}\left[\frac{1}{2K_{0}\Re z_{q}}\left(1+K_{0}^{2}|z_{q}|^{2}\right)-1\right],\label{average}\\
&&\langle\Phi(x)\Phi(x')\rangle=\frac{1}{L}\sum_{q>0}\frac{\cos\left[q(x-x')\right]}{q\Re z_{q}}.\label{eq:corr}
\end{eqnarray}

\section{Energy fluctuations}

We now describe how to compute different moments of the energy moments
using the wave-function approach introduced above. First, we present, in Sec.~\ref{subsec:pert}, a perturbative treatment of the Langevin equation \eqref{eq:lang},
valid for small deviation from the initial values, i.e. $z_{q}(t)\approx K_{0}^{-1}$.
In Sec. \ref{subsect:FK}, we then present a treatment valid for large deviation from the
initial value based on the Fokker-Planck equation.

\subsection{Perturbative treatment}\label{subsec:pert}

As we are interested in the low-energy limit of $|\delta z_{q}(t)|\ll K_{0}^{-1}$,
where $z_{q}(t)=K_{0}^{-1}+\delta z_{q}(t)$, we can linearize Eq.~(\ref{eq:lang})
in $\delta z_{q}$ and obtain a simple linear equation:~\cite{Rahmani11}
\begin{equation}
i\:\delta\dot{z}_{q}=2q\:\left(\delta z_{q}-\delta\alpha\right),\label{eq:langevin}
\end{equation}
which admits the explicit solution $\delta z_{q}(t)=2iq\int_{0}^{t}dt^{\prime}\: e^{2iq(t^{\prime}-t)}\delta\alpha(t^{\prime})$.
Our strategy for computing the leading contribution in $\delta z_q$
of a generic noise-averaged function of $z_q$ is as follows. We expand this
function to leading order in $\delta z_{q}$, and insert the explicit
solution above (in terms of $\delta\alpha$) into the resulting expression.
The average over noise can then be done by using the Wick's theorem,
which relates $\overline{\delta\alpha(t_{1})\delta\alpha(t_{2})\dots\delta\alpha(t_{n})}$
to two-point functions~(\ref{eq:corr_noise}). [As the $n$-point
function above vanishes for odd $n$ this approach is possible only
when the leading contribution is even in $\delta z$. Luckily, this
is the case for all energy moments of interest. Note that
the expansion to the next-to-leading order in $\delta z$ is not consistent
with the linearization approximation made in Eq.~\eqref{eq:langevin}.]
Finally, we perform the required integrations over the time arguments
to find the noise average.

Let us now compute the two types of energy moments~\eqref{average2}
and \eqref{average1}. Here, we only consider the first and the second
moments (see Appendix.~\ref{sec:Higher-moments-of} for a discussion
of higher moments). As stated above, all observables, including $\langle H_{q}^{m}(K_{0})\rangle$,
can be written in terms of $z_{q}$. The expressions for $\langle H_{q}^{m}(K_{0})\rangle$
become more and more involved as $m$ increases, but if we expand
these expressions to leading order in $\delta z_{q}$, we find the
simple relationship $\langle H_{q}^{m}(K_{0})\rangle=q^{m}\,2^{m-3}\, K_{0}^{2}|\delta z_{q}|^{2}+{\cal O}(\delta z_{q}^{3})$.
Using the strategy outlined above, we then
obtain $\overline{\langle H_{q}^{m}\rangle}\approx2^{m-1}q^{m+2}K_{0}^{2}W^{2}t$
and $\overline{\langle H_{q}\rangle^{2}}\approx q^{6}K_{0}^{4}W^{4}\left[2t^{2}+\sin^{2}(2qt)/4q^{2}\right]$
(see Appendix.~\ref{sec:Higher-moments-of} for details). Clearly,
for a single mode, the energy fluctuations are significantly affected
by the quantum source at least in the limit of small excess 
energy, i.e. $\overline{\langle H_{q}^{2}\rangle}\ne\overline{\langle H_{q}\rangle^{2}}$.

So far we have considered a single mode $q$ with Hamiltonian $H_{q}^{\Re}$
(or $H_{q}^{\Im}$). We now turn to the (many-mode) LL with Hamiltonian
$H=2\sum_{q>0}H_{q}$ where the factor of $2$ accounts for the contributions
of $H_{q}^{\Re}$ and $H_{q}^{\Im}$. The average energy can be simply
written as $\overline{\langle\epsilon\rangle}=2\sum_{q>0}\overline{\langle H_{q}\rangle}$,
and using $\sum_{q>0}\rightarrow\frac{L}{2\pi}\int_{0}^{\pi}$, we
then obtain 
\begin{equation}
\overline{\langle\epsilon\rangle}\approx L\pi^{3}\: K_{0}^{2}W^{2}t/4.\label{eq:E}
\end{equation}
 As for the second moment, we have $\langle\epsilon^{2}\rangle\equiv4\sum_{q_{1},q_{2}>0}\langle H_{q_{1}}H_{q_{2}}\rangle$
and $\langle\epsilon\rangle^{2}\equiv4\sum_{q_{1},q_{2}>0}\langle H_{q_{1}}\rangle\langle H_{q_{2}}\rangle$.
Noting that the many-mode wave function is a direct product
of wave functions for different modes $q$ (in the $\Re$ and $\Im$
sectors), we can then write $\overline{\langle\epsilon^{2}\rangle}=\overline{\langle\epsilon\rangle^{2}}+4\sum_{q}\left(\overline{\langle H_{q}^{2}\rangle}-\overline{\langle H_{q}\rangle^{2}}\right)$.
In the previous expression, both $\overline{\langle\epsilon\rangle^{2}}$
and $\overline{\langle\epsilon^{2}\rangle}$ scale as $L^{2}$, while
the sum over $q$ scales as $L$. Therefore, if we take the thermodynamic
limit before any other limit, the two types of moments will be, to
leading order, identical [see Fig.~\ref{fig1}(b)]. As we will see
below, this is also the case for the two types of cumulants obtained
by subtracting $\left[\overline{\langle\epsilon\rangle}\right]^{2}$
from the moments above. Such subtraction does not change the scaling
with $L^{2}$ as an explicit calculation gives (see Appendix \ref{sec:Derivation-of-Eq.}):
\begin{equation}
\overline{\langle\epsilon\rangle^{2}}-\left[\overline{\langle\epsilon\rangle}\right]^{2}=\frac{1}{16}\pi^{6}K_{0}^{4}W^{4}t^{2}L^{2}{\cal F}(\pi t),\label{eq:F}
\end{equation}
where the function ${\cal F}(x)$ has the following asymptotic behavior:
${\cal F}(x)\approx2\left(1-\frac{4}{9}x^{2}\right)$ for $x\ll1$,
and ${\cal F}(x)\approx\frac{16\pi}{7x}$ for $x\gg1$. This indicates
that the cumulant above crosses over from quadratic growth in $t$
for short times to linear growth in $t$ at longer times. Combining
Eqs.~\eqref{eq:F} and \eqref{eq:E} leads to the important relationship~\eqref{eq:main1}.
The regime of validity for these results can be extended by decreasing
the strength of the noise $W$. A comment is in order on the scaling
of Eq.~(\ref{eq:F}) with $L^{2}$. In equilibrium and for short-range interactions, all cumulants
of energy are expected to scale linearly with the system size.
The scaling of Eq. \eqref{eq:F} with $L^{2}$ is a nonequilibrium
feature, which indicates that the fluctuation of the absorbed energy
from one noise realization to the other is extensive.

\subsection{Fokker-Planck approach}\label{subsect:FK}

Our results so far have been obtained by doing perturbative calculations
in the limit of small $\delta z_{q}$. This limit coincides with the
regime of validity of the LL description and is of great experimental
interest. From a theoretical perspective, however, it is interesting
to study the effects of nonlinearities in Eq.~\eqref{eq:lang}. An
alternative approach, which allows us to go beyond the limit of small
$\delta z_{q}$, is through the Fokker-Planck (FP) equation for the
wave-function probability distribution. Using such FP equation, we
obtain below an exact nonperturbative expression for the average energy
of the system at time $t$. Note that the FP approach is valid in the white-noise limit. 

Let us briefly review the general formalism: \cite{Risken} for a vector
$\vec{a}$ of stochastic variables satisfying the Langevin equation
$\partial_{t}a_{i}=h_{i}(\vec{a})+g_{i}(\vec{a})\gamma(t)$, where
$h_{i}$ and $g_{i}$ are arbitrary functions of $\vec{a}$ and $\overline{\gamma(t)\gamma(t^{\prime})}=2\delta(t-t^{\prime})$,
the probability distribution $f(\vec{a},t)$ evolves according to
the FP equation $\partial_{t}f={\cal D}f$, with the differential
operator ${\cal D}=-\frac{\partial}{\partial a_{i}}h_{i}-\frac{\partial}{\partial a_{i}}\frac{\partial g_{i}}{\partial a_{j}}g_{j}+\frac{\partial}{\partial a_{i}}\frac{\partial}{\partial a_{j}}g_{i}g_{j}$
(summation over repeated indices is implied). The noise-averaged expectation
value of an arbitrary function $G(\vec{a})$ of the stochastic variables
$\vec{a}$ can then be computed at time $t$ as an integral over $\prod_{i}\: d\vec{a}_{i}$
weighted by the formal solution of the FP equation, $f(\vec{a},t)=e^{{\cal D}t}f(\vec{a},0)$.

In analogy with the Heisenberg picture of quantum dynamics, we can
evolve the observable $G(\vec{a})$ instead of the distribution function
$f(\vec{a})$ (using repeated integration by parts), and write
\begin{equation}
\left.\overline{G(\vec{a})}\right|_{t}=\int\prod_{i}\: d\vec{a}_{i}\: f(\vec{a},0)\: e^{{\cal D^{\dagger}}t}G(\vec{a}),\label{eq:formal2}
\end{equation}
where ${\cal D}^{\dagger}=\left(h_{i}+\frac{\partial g_{i}}{\partial a_{j}}g_{j}\right)\frac{\partial}{\partial a_{i}}+g_{i}g_{j}\frac{\partial}{\partial a_{i}}\frac{\partial}{\partial a_{j}}$.
By expanding the exponential operator above as $e^{{\cal D^{\dagger}}t}=\sum_{n}\frac{t^{n}}{n!}{\cal D^{\dagger}}^{n}$,
we can then compute the noise average of $G(\vec{a})$ as a Taylor
expansion in $t$. Since we are interested in finite time scales,
such expansion is indeed very useful even when truncated at a finite
order. For the average energy $\overline{\langle\epsilon\rangle}$ and the current-current correlation function [see Eq.~\eqref{eq:C2}], it turns
out that the Taylor series can be resummed, which results in an exact
solution (see Appendix \ref{sec:Fokker-Plank-approach-for}). In general, the short-time expansion can potentially break down after a critical time (indicating a dynamical phase transition) due to nonanalytic behavior in the Loschmidt echo or certain observables.~\cite{Wei2012,Mitra2012b,Heyl2013,Karrasch2013} However, in this case, we do not observe any nonanalyticity in the thermodynamic limit for the observables that we compute.

Let us now apply the FP approach to the average energy. As the energy
is just a sum of the single-mode energies, we need a FP equation for
a single mode, i.e., $\vec{a}=(\Re z_{q},\Im z_{q})$. We can then
explicitly construct the differential operator ${\cal D}^{\dagger}$
corresponding to Eq.~(\ref{eq:lang}) and find the $n$-th order
term by applying it $n$ times to the exact expression in terms of $\vec{a}$ [Eq. \eqref{average}] for $\langle H_{q}\rangle$. Note that
the integration over $a_{i}$ in Eq.~\eqref{eq:formal2} is trivial
due to a $\delta$-function initial distribution $f(\vec{a},0)=\delta(a_{1}-K_{0}^{-1})\delta(a_{2})$.
By explicitly computing the terms in such an expansion to order $n=10$,
we found that the series matches a simple exponential, $\overline{\langle H_{q}\rangle}=\frac{q}{2}\left[\exp\left(2q^{2}W^{2}K_{0}^{2}t\right)-1\right]$,
term by term. Modeling Eq.~\eqref{eq:langevin} by a sequence of random
quenches,~\cite{Rahmani11,Perfetto} we have also simulated the Langevin dynamics and verified this expression
numerically. Upon integration over $q$, we then obtain 
\begin{equation}
\overline{\langle\epsilon\rangle}=\frac{L}{8\pi K_{0}^{2}W^{2}t}\left(e^{2K_{0}^{2}\pi^{2}W^{2}t}-2\pi^{2}K_{0}^{2}W^{2}t-1\right).
\end{equation}
 As expected, the above expression, up to first order in time, agrees
with Eq.~\eqref{eq:E}.

Using the Wigner-function approach~\cite{Wigner_paper} (see also Ref.~\onlinecite{Polkovnikov}),
we can additionally show that for a thermal initial state (which is
subsequently decoupled from the thermal environment during the evolution),
the absorbed single-mode energy above is simply multiplied by a prefactor
$\coth\left(q/2k_{B}T_{0}\right)$, where $T_{0}$ is the initial
temperature, and $k_{B}$ the Boltzmann constant~(see Appendix \ref{sec:Wigner-function-approach}).
The many-mode energy can then be similarly computed by integration
over $q$. The FP approach can be used to compute other quantities
such as the second moment of energy. However, in this case, one needs
to construct a FP equation for four stochastic variables, $\vec{a}=(\Re z_{q_{1}},\Im z_{q_{1}},\Re z_{q_{2}},\Im z_{q_{2}})$,
and so on and so forth for higher moments.

\section{Correlation Functions}

In this section, we compute some noise-averaged correlation
functions of the system. We start by the current-current correlation function (see Appendix \ref{sec:Expression-for-observables}):
\begin{equation}
C(x-x^{\prime})\equiv\langle\partial_{x}\Phi(x)\:\partial_{x'}\Phi(x')\rangle=\frac{1}{L}\sum_{q>0}\frac{q\,\cos\left[q(x-x')\right]}{\Re z_{q}}.\label{eq:C2}
\end{equation}
Using the perturbative method of Sec.~\ref{subsec:pert}, we can expand the above expression in $\delta z_q=z_q-K_0^{-1}$ as 
\begin{equation}\label{eq:corr2_expand}
{\cal C}(x)= {K_0  \over L}\sum_{q>0}
\:q\;{\cos(q x)}\left[1-K_0(\Re \delta z_q)+K_0^2{(\Re \delta z_q)^2}+\cdots \right].
\end{equation}
If we now use the solution of the linearized Langevin equation~\eqref{eq:langevin}, we find that the first-order term in $\delta z$ vanishes upon noise averaging, and the second-order term gives a leading contribution of order $W^2$:
\begin{equation}\label{eq:second}
\overline{(\Re\delta z_{q})^{2}}\approx\frac{W^{2}}{2}q\left[4qt-\sin(4qt)\right].
\end{equation}
Due to the presence of a linear term in Eq.~\eqref{eq:corr2_expand}, however, we need to also expand the Langevin equation~\eqref{eq:lang} to second order in $\delta z$:
\begin{equation}\label{eq: dddd}
i\:\delta\dot{z}_{q}=2q\:\left(\delta z_{q}-\delta\alpha\right)+K_0q \delta z_q^2 -2K_0 q\delta z_q \delta \alpha,
\end{equation}
which may lead to a contribution of order $W^2$ to $\overline{\Re \delta z_q}$. We then proceed by computing the next correction to $\delta z$ iteratively: we write $\delta z=\delta z^{(1)}+\delta z^{(2)}$, where $\delta z^{(1)}$ is first-order in $\delta \alpha$ and satisfies the linear equation~\eqref{eq:langevin} [it is given by the explicit integral expression below Eq. ~\eqref{eq:langevin}]. We then insert the above $\delta z$ into Eq.~\eqref{eq: dddd} to obtain an equation for the evolution of $\delta z^{(2)}$:
\begin{equation}
i \delta \dot{z}_q^{(2)}=2q\left[\delta z_q^{(2)}-K_0\delta z^{(1)}\delta \alpha+K_0(\delta z^{(1)})^2/2\right],
\end{equation}
where we have kept only the second-order terms in $\delta \alpha^2$.
The above linear equation for $\delta z^{(2)}$ yields an explicit expression in terms of $\delta z^{(1)}$ and $\delta \alpha$. Replacing $\delta z^{(1)}$ by the explicit solution of Eq.~\eqref{eq:langevin} leads to an integral expression for $\delta {z}_q^{(2)}$ in terms of $\delta \alpha$. After some algebra, we can then write the leading contribution to $\overline{\Re \delta z_q}$ as
\begin{equation}\label{eq:first}
\overline{\Re \delta z_q(t)}\approx -{K_0  W^2 \over 2} q\sin (4qt),
\end{equation}
which is of the same order in $W$ as Eq.~\eqref{eq:second}.
Inserting Eqs.~\eqref{eq:first} and \eqref{eq:second} into Eq.~\eqref{eq:corr2_expand}, and performing an integral over $q$ gives
\begin{equation}\label{eq:sh}
\overline{{\cal C}(x)}\approx -{K_0 \over 2 \pi}\left({1 \over x^2}+12 K_0^2W^2 {t \over x^4} \right),
\end{equation}
where we have neglected fast oscillatory terms for asymptotically large $x$.

We now turn to the correlation functions of vertex operators:
\begin{equation}
{\cal V}(x-x')\equiv\langle e^{i\nu\Phi(x)}\: e^{-i\nu\Phi(x')}\rangle=\exp\left[-\frac{\nu^{2}}{2}\langle\Phi(x)\Phi(x')\rangle\right],\label{eq:C3}
\end{equation}
where $\nu$ is a generic constant (which depends on $K_0$ for realistic models), and the correlator in the argument of the exponential is given by Eq.~\eqref{eq:corr}. As in Eq.~\eqref{eq:corr2_expand}, we expand the above expression up to second order in $\delta z_q$ as $
{{\cal V}(x)}\approx{e^{-\frac{\nu^{2}}{2L}\sum_{q>0}K_{0}\frac{\cos(qx)}{q}\left(1-K_{0}\:{(\Re\delta z_{q})}+K_{0}^{2}\:{(\Re\delta z_{q})^{2}}\right)}}$,
 which gives \begin{widetext} 
 \begin{equation}
\label{long-eq}
{{\cal V}(x)} \approx e^{-\frac{\nu^{2}}{2L}\sum_{q>0}K_{0}\frac{\cos(qx)}{q}}\left[1+\frac{\nu^{2}}{2L}K_{0}^{2}\sum_{q>0}\frac{\cos(qx)}{q}\:\left((\Re\delta z_{q})-K_0{(\Re\delta z_{q})^{2}}\right)+\frac{\nu^{4}}{8L^{2}}K_{0}^{4}\sum_{q_{1},q_{2}>0}\frac{\cos(q_{1}x)\cos(q_{2}x)}{q_{1}q_{2}}{(\Re\delta z_{q_{1}})(\Re\delta z_{q_{2}})}\right].
\end{equation}
\end{widetext} 
To perform the averaging over noise we need Eqs.~\eqref{eq:second} and \eqref{eq:first} as well as $\overline{(\Re\delta z_{q_{1}})(\Re\delta z_{q_{2}})}$, which can be computed from the solution of the linearized Langevin equation and gives:
\begin{equation}
\label{two-modes}
\begin{split}\overline{(\Re\delta z_{q_{1}})(\Re\delta z_{q_{2}})}&=q_{1}q_{2}W^{2}
\\&\times\left(\frac{\sin[2(q_{1}-q_{2})t]}{q_{1}-q_{2}}-\frac{\sin[2(q_{1}+q_{2})t]}{q_{1}+q_{2}}\right).\end{split}
\end{equation}
Note that the above expression reduces to Eq.~\eqref{eq:second} in the limit of $q_1 \rightarrow q_2$.

If we now take the asymptotic limit of large $x$ in Eq.~\eqref{long-eq}, we find that the main contributions to the integrals over momenta (we are using $\sum_{q>0}\rightarrow\frac{L}{2\pi}\int_{0}^{\pi}$) come from small $q$. For fixed time $t \ll x$, we can then expand $\overline{(\Re\delta z_{q_{1}})(\Re\delta z_{q_{2}})}\approx {16\over3} W^2 t^3 q_1^2 q_2^2$. 
Upon integration over momenta (and neglecting the fast oscillations) the term proportional to $\nu^2$ in \eqref{long-eq} gives a contribution scaling as $t/x^2$, while the term proportional to $\nu^4$ gives a subleading contribution scaling as $t^3/x^4$. We then find that the leading correction (in the limit $t \ll x$) to the ground-state correlation function is ${\cal V}(x)\approx{\cal V}^{(0)}(x) (1+\delta{\cal V}(x))$ 
with $\delta{\cal V}(x)\propto \nu^2 K_0^3 W^2 t x^{-2}$.

Interestingly, for the current-current correlation function~\eqref{eq:C2}, the FP approach can provide an exact solution. Since this correlation function is the sum of contributions for each
momentum mode, we can use the FP approach for each modes separately.
In this case the variable of interest is $\left(\Re z_{q}\right)^{-1}$ [see Eq.~\eqref{eq:C2}]. 
By repeating the procedure used to compute the excess energy (see
Appendix \ref{sec:Fokker-Plank-approach-for}), we obtain a Taylor series
that can be resummed to give the exact expression 
\begin{equation}
\overline{\left(\Re z_{q}\right)^{-1}}=K_{0}\left[\exp\left(2K_{0}^{2}W^{2}q^{2}t\right)\right]\label{eq:one_over_R}
\end{equation}

Plugging the expression (\ref{eq:one_over_R}) in Eq.~(\ref{eq:C2}) results in the exact integral expression for $\overline{{\cal C}(x)}$, which can be written in closed form in terms of the error function:
\begin{equation}
\overline{{\cal C}(x)}=\frac{K_{0}}{2\pi}\int_{0}^{\pi}dq\, q\,\cos\left(q\, x\right)\exp\left(2K_{0}^{2}W^{2}q^{2}t\right),\label{eq:C2_result}
\end{equation}
which agrees with Eq.~\eqref{eq:sh} in the limit of small absorbed energy.

The noise-averaged current-current correlation function above is related, by double differentiation, 
to the noise-averaged bosonic correlation function (averaging over noise commutes with differentiation). On the other hand, the simple relationship~\eqref{eq:C3} between the vertex operator and the bosonic correlation functions does not hold for noise-averaged correlators [as seen in  Eq.~\eqref{two-modes} averaging over noise does not commute with exponentiation due to the correlations between different modes].

\section{Conclusion}

In summary, we studied a thermally isolated LL, randomly driven with
a noisy Luttinger parameter, and undergoing coherent quantum evolution
for each realization of noise. We computed noise-averaged correlation
functions, and studied the energy fluctuations. We characterized these
fluctuations by two types of energy moments: one that mixes the classical
and quantum sources of fluctuations and one that is only affected
by the classical source. We found that while for a single mode, the
two types of moments lead to very different results, for the many-mode
problem, the difference disappears in the thermodynamic limit. This
indicates that many-body properties of such noise-driven coherent
systems likely exhibit effective decoherence. Our approach to the
dynamics of noisy LLs is based on a mapping of the quantum problem
to one of nonequilibrium classical statistical mechanics, which provides
powerful tools, such as the FP equation, for performing exact calculations.

\acknowledgements We are grateful to K. Barros, C. Chamon, A. del
Campo, T. Giamarchi, D. Huse, P. Krapivsky, I. Martin, A. Polkovnikov,
P. Zoller and W. Zurek for helpful discussions. This work was supported
in part by AFOSR FA9550-10-1-0110 (LD) and the U.S. Department of
Energy through LANL/LDRD program (AR).

\appendix

\section{EIGENSTATE EXPANSION OF THE
SINGLE-MODE WAVE FUNCTION\label{sec:Eigenstate-expansion-of}}

Here we compute the overlaps of the single-mode wave function, Eq.~(\ref{eq:psi}),
with the eigenstates of the single-mode Hamiltonian $H_{q}$. The
Hamiltonian $H_{q}$ is shorthand for either $H_{q}^{\Re}$ or $H_{q}^{\Im}$:
\begin{equation}
H_{q}(K(t))=u\left(-\frac{K(t)}{4}\:\partial_{\phi}^{2}+\frac{1}{K(t)}\: q^{2}\phi^{2}\right).\label{eq:Hq}
\end{equation}
In the absence of noise, the Hamiltonian is given by $H_{0}\equiv H_{q}(K_{0})$.
For $z_{q}=1/K_{0}$, the wave function (\ref{eq:psi}) is the ground
state of Hamiltonian (\ref{eq:Hq}). For arbitrary $z_{q}$, however,
it is a superposition $\psi_{q}(\phi)=\sum_{n=0}^{\infty}c_{n}(z_{q})\psi_{n}(\phi)$,
where $\psi_{n}(\phi)$ is an eigenfunction of $H_{0}$ with energy
$\left(n+\frac{1}{2}\right)uq$, which can be written explicitly in
terms of the Hermite polynomials. Through direct integration, we can
compute the amplitudes $c_{n}(z_{q})=\int d\phi\:\psi_{q}(\phi)^{\star}\psi_{n}(\phi)$, which vanish for odd $n$, and are given by the following expression for even $n$:
\[
|c_{2m}(z_{q})|^{2}=\frac{(2m)!\:x_{q}^{1/2}}{2^{2m-1}m!m!}\,{\left[(x_{q}-1)^{2}+y_{q}^{2}\right]^{m}}{\left[(x_{q}+1)^{2}+y_{q}^{2}\right]^{-m-1/2}},
\]
where $x_{q}\equiv K_{0}\left(\Re\: z_{q}(t)\right)$ and $y_{q}\equiv K_{0}\left(\Im\: z_{q}(t)\right)$.
Thus, the overlaps $|c_{n}(z_{q})|$ are identically zero for odd
$n$, and decay exponentially for even $n$.

\section{EXPRESSION FOR OBSERVABLES IN TERMS
OF $z_{q}$\label{sec:Expression-for-observables}}

In this appendix, we express some of the observables of the system in terms of $z_{q}(t)$.
Measuring the energies with respect to the ground state of $H_{q}(K_{0})$,
and using the explicit form of the wave function (\ref{eq:psi}),
we can obtain Eq.~(\ref{average}) by direct integration over the
bosonic fields:
\begin{equation}
\begin{split}
\langle H_{q}(K_{0})\rangle&\equiv\int_{-\infty}^{+\infty}d\psi\:\psi_{q}(\phi,t)^{\star}H_{q}(K_{0})\psi_{q}(\phi,t)\\
&=\frac{q}{2}\left[\frac{1}{2K_{0}\Re z_{q}}\left(1+K_{0}^{2}|z_{q}|^{2}\right)-1\right],
\end{split}
\end{equation}
where $u$ is set to unity. Similarly, we can compute the second
moment of energy:
\begin{equation}
\begin{split}
\langle H_{q}^{2}(K_{0})\rangle-&\langle H_{q}(K_{0})\rangle^{2}\\
&=\frac{q^{2}}{8K^{2}\Re z_{q}^{2}}\left[1-2K_{0}^{2}(\Re z_{q}^{2}-\Im z_{q}^{2})+K_{0}^{4}|z_{q}|^{4}\right].
\end{split}
\end{equation}
Note that both expressions above vanish for the ground state ($z_{q}=K_{0}^{-1}$).
This simple approach can be used to compute higher moments of energy,
but the resulting expressions become more cumbersome. Using\begin{footnotesize} MATHEMATICA\textregistered \end{footnotesize},
we have calculated the moments $\langle H_{q}^{m}(K_{0})\rangle$
for $m\le20$, and checked that to leading order in $\delta z_{q}=z_{q}-K_{0}^{-1}$,
they can be written as
\begin{equation}
\langle\epsilon_{q}^{m}\rangle=\langle H_{q}^{m}(K_{0})\rangle=q^{m}\,2^{m-3}\, K_{0}^{2}|\delta z_{q}|^{2}+O(\delta z_{q}^{3}).\label{eq:H^m_q}
\end{equation}
Equal-time correlation functions are also simple to compute in terms
of $z_{q}(t)$. We can expand $\Phi(x)$ in Fourier modes, $\Phi(x)=\sum_{q}\frac{e^{iqx}}{\sqrt{L}}\Phi_{q}=\frac{2}{\sqrt{L}}\sum_{q>0}\left[\cos(qx)\Re\Phi_{q}-\sin(qx)\Im\Phi_{q}\right]$,
and make use of the relation $\langle\Re\Phi_{q_{1}}\Re\Phi_{q_{2}}\rangle=\langle\Im\Phi_{q_{1}}\Im\Phi_{q_{2}}\rangle=\delta_{q_{1}q_{2}}/4q\Re z_{q}$
and $\langle\Re\Phi_{q_{1}}\Im\Phi_{q_{2}}\rangle=0$ to obtain Eq.~(\ref{eq:corr}):
\begin{equation}
\langle\Phi(x)\Phi(x')\rangle=\frac{1}{L}\sum_{q>0}\frac{\cos\left[q(x-x')\right]}{q\Re z_{q}}.
\end{equation}
Other correlation functions for the current and the vertex operator
can be obtained by simple differentiation and exponentiation.

%

\section{HIGHER MOMENTS OF ENERGY\label{sec:Higher-moments-of}}

Focusing on a single mode for simplicity, we calculate the noise-averaged
moments $\overline{\langle\epsilon_{q}\rangle^{m}}$. To leading order
in $\delta z_{q}$, we can write $\langle\epsilon_{q}\rangle^{m}\approx\left(\frac{K_{0}^{2}q}{4}|\delta z_{q}|^{2}\right)^{m}$ [see Eq.~(\ref{eq:H^m_q})].
 After substituting the explicit solution of the linear Langevin equation~\eqref{eq:langevin}
we need to compute 
\begin{equation}
\begin{split} & \overline{|\delta z_{q}|^{2m}}=(2q)^{2m}\int_{0}^{t}dt_{1}e^{2iqt_{1}}\int_{0}^{t}dt_{1}^{\prime}e^{-2iqt_{1}^{\prime}}\cdots\int_{0}^{t}dt_{m}^{\prime}e^{-2iqt_{m}^{\prime}}\\
 & \times\overline{\delta\alpha(t_{1})\delta\alpha(t_{1}^{\prime})\cdots\delta\alpha(t_{m})\delta\alpha(t_{m}^{\prime})}.
\end{split}
\end{equation}
For Gaussian white noise, $\overline{\delta\alpha(t)\delta\alpha(t^{\prime})}=W^{2}\delta(t-t^{\prime})$,
the above $2m$-point function breaks up into a sum of products of
$m$ delta functions. The number of terms in this sum, i.e., the number
of different contractions, is $(2m-1)!!$. There are three different
types of contractions: (i) $\overline{\delta\alpha(t_{i})\delta\alpha(t_{j})}$,
(ii) $\overline{\delta\alpha(t_{i}^{\prime})\delta\alpha(t_{j}^{\prime})}$,
and (iii) $\overline{\delta\alpha(t_{i})\delta\alpha(t_{j}^{\prime})}$.
Note that we have an equal number of contractions of type (i) and
(ii), which are related to each other by complex conjugation. Upon
integration, we obtain $
\int dt_{i} dt_{j}^{\prime}\:e^{2iq(t_{i}-t_{j}^{\prime})}\overline{\delta\alpha(t_{i})\delta\alpha(t_{j}^{\prime})}=W^{2}t
$
 and
$
\int dt_{i} dt_{j} dt_{m}^{\prime} dt_{n}^{\prime} e^{2iq(t_{i}+t_{j})}\overline{\delta\alpha(t_{i})\delta\alpha(t_{j})}e^{-2iq(t_{m}+t_{n})}
\overline{\delta\alpha(t_{m}^{\prime})\delta\alpha(t_{n}^{\prime})}=  W^{4}\left(\frac{\sin(2qt)}{2q}\right)^{2}
$ (all integrals are from $0$ to $t$),
 which allows us to write 
\begin{equation}
\overline{\langle\epsilon_{q}\rangle^{m}}\approx q^{3m}\,(K_{0}^{2}W^{2}t)^{m}\,\sum_{p=0}^{\lfloor m/2\rfloor}S_{m,p}\left(\frac{\sin2qt}{2qt}\right)^{2p},\label{single_mode}
\end{equation}
 where $\lfloor x\rfloor$ indicates the floor value of $x$, and $S_{m,p}$
represents the number of ways for having $p$ contractions of type
(i), $p$ contractions of type (ii), and, consequently, $m-2p$ contractions
of type (iii). The simple combinatorial argument below gives 
\begin{equation}
S_{m,p}=\left[\left(\begin{array}{c}
m\\
2p
\end{array}\right)\cdot(2p-1)!!\right]^{2}(m-2p)!.\label{eq:S_combinatorics}
\end{equation}
 Out of the $m$ choices for $\delta\alpha(t_{i})$, we select $2p$,
and then make $p$ contractions among them in $(2p-1)!!$ ways. A
similar argument applies to $\delta\alpha(t_{i}^{\prime})$, which
results in the power of $2$ for the term in the square brackets.
We now have $m-2p$ unpaired $\delta\alpha(t_{i})$, and $m-2p$ unpaired
$\delta\alpha(t_{i}^{\prime})$, which can be paired up in $(m-2p)!$
ways.

\section{DERIVATION OF EQ. \eqref{eq:F} \label{sec:Derivation-of-Eq.}}

To compute ${\cal F}(x)$ in Eq. (\ref{eq:F}), we start from
the expression $\langle\epsilon\rangle=2\sum_{q}\langle H_{q}\rangle$
(the factor of $2$ accounts for the contribution of $H_{q}^{\Re}$
and $H_{q}^{\Im}$), and write $\overline{\langle\epsilon\rangle^{2}}=4\sum_{q_{1},q_{2}}\overline{\langle H_{q_{1}}\rangle\langle H_{q_{2}}\rangle}$.
After inserting Eq. \eqref{eq:H^m_q} for $m=1$, we obtain 
\begin{equation}
\begin{split}\overline{\langle\epsilon\rangle^{2}}\approx & \sum_{q_{1},q_{2}}\,\frac{K_{0}^{4}q_{1}q_{2}}{4}\,\overline{|\delta z_{q_{1}}|^{2}|\delta z_{q_{2}}|^{2}},\end{split}
\end{equation}
 which, upon substituting the explicit solution of Eq.~\eqref{eq:langevin}, yields 
\begin{equation}
\begin{split}\overline{\langle\epsilon\rangle^{2}}\approx4K_{0}^{4}\sum_{q_{1},q_{2}}q_{1}^{3}q_{2}^{3}\iiiint_{0}^{t} & \,\, dt_{1}dt_{2}dt_{1}^{\prime}dt_{2}^{\prime}\,\, e^{2iq_{1}(t_{1}-t_{1}^{\prime})}e^{2iq_{2}(t_{2}-t_{2}^{\prime})}\\
 & \times\overline{\delta\alpha(t_{1})\delta\alpha(t_{1}^{\prime})\delta\alpha(t_{2})\delta\alpha(t_{2}^{\prime})}.
\end{split}
\end{equation}
Using the Wick's theorem to break the four-point function $\overline{\delta\alpha(t_{1})\delta\alpha(t_{1}^{\prime})\delta\alpha(t_{2})\delta\alpha(t_{2}^{\prime})}$
into a sum of products of two-point functions, we get three different
types of contractions. Upon inserting these delta functions and performing
the integrals over time, we then find 
\begin{equation}
\begin{split}\overline{\langle\epsilon\rangle^{2}}=4K_{0}^{4}W^{4} & \sum_{q_{1},q_{2}}q_{1}^{3}q_{2}^{3}\\
 & \times\left\{ \frac{\sin^{2}\left[(q_{1}+q_{2})t\right]}{(q_{1}+q_{2})^{2}}+\frac{\sin^{2}\left[(q_{1}-q_{2})t\right]}{(q_{1}-q_{2})^{2}}+t^{2}\right\} .
\end{split}
\label{corr}
\end{equation}
 The term proportional to $t^{2}$ is exactly equal to $\left[\overline{\langle\epsilon\rangle}\right]^{2}$.
To obtain the asymptotic value of ${\cal F}(x)$, we convert the sums
to integrals, and define the new variables $q_{i}=\pi\xi_{i}$ and
$x=\pi t$, which leads to

\begin{equation}
\begin{split}
{\cal F}(x)= & 16\int_{0}^{1}d\xi_{1}\int_{0}^{1}d\xi_{2}\:\xi_{1}^{3}\xi_{2}^{3}\\
 & \times\left\{ \frac{\sin^{2}\left[(\xi_{1}+\xi_{2})x\right]}{\left[(\xi_{1}+\xi_{2})x\right]^{2}}+\frac{\sin^{2}\left[(\xi_{1}-\xi_{2})x\right]}{\left[(\xi_{1}-\xi_{2})x\right]^{2}}\right\}. 
\end{split}\label{eq:F_int}
\end{equation}
For $x\ll1$, the asymptotic behavior is obtained by a simple Taylor
expansion in $x$, which gives ${\cal F}(x)\sim2\left(1-\frac{4}{9}x^{2}\right)$.
For $x\gg1$, the main contribution comes from $\xi_{1}\approx\xi_{2}$;
then we can approximate ${\cal F}(x)\sim16\,\frac{1}{2}\,\int dQ\int dq\left(\frac{Q+q}{2}\right)^{3}\left(\frac{Q-q}{2}\right)^{3}\frac{\sin^{2}(qx)}{(qx)^{2}}$
where we have defined $Q=\xi_{1}+\xi_{2}$ and $q=\xi_{1}-\xi_{2}$,
and the factor $1/2$ comes from the Jacobian of the transformation.
Since $Q\gg q$ the dominant contribution comes from the term proportional
to $Q^{6}$. The final change of variable $k=qx$ leads to ${\cal F}(x)\sim\frac{16}{2^{7}}\int_{0}^{2}dQ\int_{-\infty}^{+\infty}\frac{dk}{x}\,\,\frac{Q^{6}\,\sin^{2}(k)}{k^{2}}=\frac{16\pi}{7x}$.

\section{EFFECTS OF COLORED NOISE\label{sec:Effects-of-colored}}

In this appendix, we discuss the effects of finite correlation time
$\tau$ in the Ornstein-Uhlenbeck process: 
\begin{equation}
\overline{\delta\alpha(t_{1})\delta\alpha(t_{1}^{\prime})}=\frac{W^{2}}{2\tau}e^{-|t_{1}-t_{1}^{\prime}|/\tau}.\label{OH}
\end{equation}
 For the single-mode energy moments $\overline{\langle\epsilon_{q}^{m}\rangle}\approx q^{m}2^{m-3}K_{0}^{2}\,\,\overline{|\delta z_{q}|^{2}}$,
we need to evaluate 
\begin{equation}
\overline{|\delta z_{q}|^{2}}=(2q)^{2}\int_{0}^{t}dt_{1}\int_{0}^{t}dt_{1}^{\prime}\, e^{2iq(t_{1}-t_{1}^{\prime})}\,\overline{\delta\alpha(t_{1})\delta\alpha(t_{1}^{\prime})}.
\end{equation}
 Substituting Eq.~\eqref{OH}, and performing the integral yields
\begin{equation}
\overline{(|\delta z_{q}|^{2})}=4q^{2}W^{2}t\left[\frac{1}{1+(2q\tau)^{2}}+\frac{\tau}{t}\Re\left(\frac{e^{-\frac{t}{\tau}(1+2q\tau)}-1}{(1+2iq\tau)^{2}}\right)\right].\label{dz2}
\end{equation}
 Note that, for $t\gg\tau$, the above expression simplifies to $\overline{|\delta z_{q}|^{2}}=4q^{2}\frac{W^{2}}{1+(2q\tau)^{2}}\, t$,
while for white noise, $\tau\rightarrow0^{+}$, we obtain $\overline{|\delta z_{q}|^{2}}=4q^{2}W^{2}t$.
We then conclude that, for $t\gg\tau$, the single-mode energy can
be obtained from the corresponding expression for white noise by the
simple rescaling $W^{2}\rightarrow\frac{W^{2}}{1+(2q\tau)^{2}}$.
The average energy can then be calculated by integration over momentum:
\begin{equation}
\begin{split}\overline{\langle\epsilon\rangle} & =2\sum_{q}\overline{\langle\epsilon_{q}\rangle}\approx K_{0}^{2}W^{2}t\frac{L}{\pi}\int_{0}^{\pi}dq\frac{q^{3}}{1+(2q\tau)^{2}}\\
 & =\frac{L}{\pi}\,\,\frac{K_{0}^{2}W^{2}t}{32\tau^{4}}\,\,\left\{(2\pi\tau)^{2}-\ln\left[1+(2\pi\tau)^{2}\right]\right\},
\end{split}
\end{equation}
 which for white noise, $\tau\rightarrow0^{+}$, reduces to $\overline{\langle\epsilon\rangle}=L\pi^{3}K_{0}^{2}W^{2}t/4$.

Let us now consider the higher moments of the single-mode energy,
which are given by $\langle\epsilon_{q}\rangle^{m}\approx\left(\frac{K_{0}^{2}q}{4}|\delta z_{q}|^{2}\right)^{m}$. Using the method of Appendix.~\ref{sec:Higher-moments-of}, and taking the limit of  $t\gg\tau$, we  can write the analog of Eq.~\eqref{single_mode}
as 
\begin{equation}
\begin{split}\overline{\langle\epsilon_{q}\rangle^{m}}\approx & \left(q^{3}K_{0}^{2}\frac{W^{2}}{1+(2q\tau)^{2}}t\right)^{m}\\
 & \times\sum_{p=0}^{\lfloor m/2\rfloor}S_{m,p}\left(\frac{\sin(2qt)-2q\tau\cos(2qt)}{2qt}\right)^{2p},
\end{split}
\end{equation}
 where $S_{m,p}$ is defined in Eq.~(\ref{eq:S_combinatorics}). We thus observe that finite correlation time does not qualitatively change the white-noise behavior.
%
Finally using the same method as in Appendix \ref{sec:Derivation-of-Eq.},
we can generalize Eq. \eqref{eq:F_int} to 
\begin{equation}
\overline{\langle\epsilon\rangle^{2}}=\left[\overline{\langle\epsilon\rangle}\right]^{2}+\frac{1}{16}L^{2}K_{0}^{4}W^{4}t^{2}\pi^{6}{\cal F}(\pi t,\pi\tau),
\end{equation}
 where, for $t\gg\tau$, we have \begin{widetext} 
\begin{equation}
\begin{split}{\cal F}(x,y)\equiv 16\int_{0}^{1}d\xi_{1}\int_{0}^{1}d\xi_{2}\,\,\xi_{1}^{3}\xi_{2}^{3}\Bigg[ & \left(\frac{(\xi_{1}+\xi_{2})y(1+4\xi_{1}\xi_{2}y^{2})\cos\left[(\xi_{1}+\xi_{2})x\right]-[1+2(\xi_{1}^{2}y^{2}+\xi_{2}^{2}y^{2})]\sin\left[(\xi_{1}+\xi_{2})x\right]}{x(\xi_{1}+\xi_{2})[1+(2\xi_{1}y)^{2}][1+(2\xi_{2}y)^{2}]}\right)^{2}\\
 & +\left(\frac{(\xi_{1}-\xi_{2})y(1-4\xi_{1}\xi_{2}y^{2})\cos\left[(\xi_{1}-\xi_{2})x\right]-[1+2(\xi_{1}^{2}y^{2}+\xi_{2}^{2}y^{2}])\sin\left[(\xi_{1}-\xi_{2})x\right]}{x(\xi_{1}-\xi_{2})[1+(2\xi_{1}y)^{2}][1+(2\xi_{2}y)^{2}]}\right)^{2}\Bigg].
\end{split}
\label{Fxy_eq}
\end{equation}
 \end{widetext} In writing the above expression, we have defined $q_{i}=\pi\xi_{i},\,\, x=\pi t,\,\, y=\pi\tau$.
Note that this expression is valid only for $t\gg\tau$, which implies
$x\gg y$. The asymptotic behavior at long times, $x\gg1$, is given
by ${\cal F}(x,y)\simeq\frac{16\pi}{7x}\left(1-\frac{56}{9}y^{2}+\mathcal{O}(y^{4})\right)$.
This implies that the finite correlation time ($y>0$) simply lowers,
by an overall prefactor, the white-noise ($y=0$) result.

\begin{figure}
\includegraphics[width=7cm]{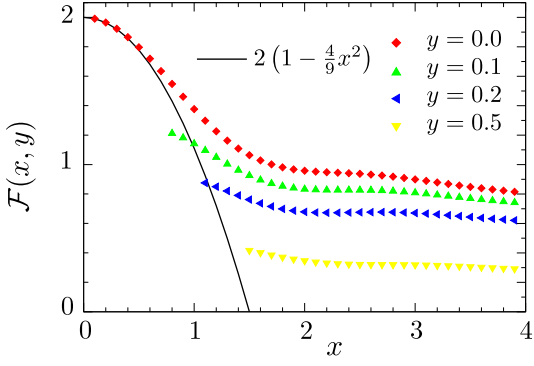} \caption{(Color online) Equation~\eqref{Fxy_eq} for $y=\pi\tau=0,\,0.1,\,0.2,\,0.5$. Note that
the equation is valid for $x\gg y$. The $y=0$ line corresponds to
the white-noise case. The asymptotic behavior for $x\ll1$ and $y=0$
is also shown (solid black line).}

\label{Fxy} 
\end{figure}

The short-time behavior can be seen in Fig.~\ref{Fxy} for several
values of $y=\pi\tau$. In the physically relevant region, $x\gg y$,
the curves show a rapid decay followed by a slower one. This behavior
is qualitatively similar to the white-noise case, which is also plotted
in Fig.~\ref{Fxy} for comparison. 

\section{FOKKER-PLANCK APPROACH FOR THE EXCESS ENERGY \label{sec:Fokker-Plank-approach-for}}

In this section we explicitly construct the differential operator
${\cal D^{\dagger}}$ for the single mode FP equation and we use it
to compute exactly the noise averaged absorbed energy.

Defining $\mathscr{R}_{i}\equiv\Re z_{q_{i}}$, and $\mathscr{I}_{i}\equiv\Im z_{q_{i}}$,
we can write the nonlinear Langevin equation~\eqref{eq:lang} as
\begin{equation}
\begin{split} & \dot{\mathscr{R}}_{i}=2K_{0}q_{i}\:\mathscr{R}_{i}\mathscr{I}_{i}-2K_{0}^{2}q_{i}\:\mathscr{R}_{i}\mathscr{I}_{i}\:\delta\alpha,\\
 & \dot{\mathscr{I}}_{i}=K_{0}q_{i}\:\left(\mathscr{I}_{i}^{2}-\mathscr{R}_{i}^{2}+K_{0}^{-2}\right)-K_{0}^{2}q_{i}\:\left(\mathscr{I}_{i}^{2}-\mathscr{R}_{i}^{2}-K_{0}^{-2}\right)\:\delta\alpha.
\end{split}
\label{nl_lang}
\end{equation}
As the total energy is a sum of single-mode energies, we only need
a single-mode Fokker-Planck equation, and we can drop the subscript
$i$. Rescaling $\delta\alpha\rightarrow\frac{W}{\sqrt{2}}\gamma(t)$,
with $\langle\gamma(t)\gamma(t^{\prime})\rangle=2\delta(t-t^{\prime})$,
we can read off $f_{i}$ and $g_{i}$ for $i=1,2$ [see discussion
after Eq.~\eqref{eq:formal2}]: 
\begin{eqnarray*}
f_{1} & = & 2K_{0}qa_{1}a_{2},\\
g_{1} & = & -\sqrt{2}WK_{0}^{2}qa_{1}a_{2},\\
f_{2} & = & K_{0}q\left(a_{2}^{2}-a_{1}^{2}+K_{0}^{-2}\right),\\
g_{2} & = & -\frac{W}{\sqrt{2}}K_{0}^{2}q\left(a_{2}^{2}-a_{1}^{2}-K_{0}^{-2}\right),
\end{eqnarray*}
 where $\vec{a}\equiv(a_{1},a_{2})=(\mathscr{R},\mathscr{I})$. In
terms of the above $f_{i}$ and $g_{i}$, we can write 
\begin{equation}
{\cal D}^{\dagger}=D_{1}^{(1)}\partial_{a_{1}}+D_{2}^{(1)}\partial_{a_{2}}+D_{1,1}^{(2)}\partial_{a_{1}}^{2}+D_{2,2}^{(2)}\partial_{a_{2}}^{2}+2D_{1,2}^{(2)}\partial_{a_{1}}\partial_{a_{2}},\label{eq:Dstar}
\end{equation}
 where 
\begin{eqnarray*}
D_{1}^{(1)} & = & f_{1}+g_{1}\frac{\partial g_{1}}{\partial a_{1}}+g_{2}\frac{\partial g_{1}}{\partial a_{2}},\\
D_{2}^{(1)} & = & f_{2}+g_{1}\frac{\partial g_{2}}{\partial a_{1}}+g_{2}\frac{\partial g_{2}}{\partial a_{2}},\\
D_{i,j}^{(2)} & = & g_{i}g_{j}.
\end{eqnarray*}
(no sum is implied). By applying Eq.~(\ref{eq:Dstar}) repeatedly
on the expression for the mean energy (see Eq. (\ref{average})):

\begin{equation}
\begin{split}\langle H_{q}\rangle=\frac{q}{2}\left[\frac{1}{2K_{0}a_{1}}\left(1+K_{0}^{2}\left(a_{1}^{2}+a_{2}^{2}\right)\right)-1\right]\end{split}
\label{eq:average-FP}
\end{equation}
and evaluating the terms at $a_{1}=K_{0}^{-1}$ and $a_{2}=0$ we
obtain a Taylor expansion which can be resummed to give the Equation $\overline{\langle H_{q}\rangle}=\frac{q}{2}\left[\exp\left(2q^{2}W^{2}K_{0}^{2}t\right)-1\right]$.
The total noise averaged excess energy is obtained by summing the
previous expression over the momentum.  

Note that if we are interested in two-mode quantities, we need to repeat
the procedure above with $\vec{a}=(\mathscr{R}_{1},\mathscr{I}_{1},\mathscr{R}_{2},\mathscr{I}_{2})$,
identify $f_{i},g_{i}$ for $i=1,2,3,4$, and build the four-dimensional
vector $D^{(1)}$ and the $4\times4$ matrix $D^{(2)}$. In general,
if we need the correlations of $m$ momenta, $D^{(1)}$ is $2m$-dimensional
and $D^{(2)}$ is a $2m\times2m$ matrix.

\section{WIGNER-FUNCTION APPROACH\label{sec:Wigner-function-approach}}

An alternative approach to the dynamics of the problem is through
the Wigner-function \cite{Wigner_paper} representation. This approach allows us to treat,
on the same footing, initial conditions given by a finite-temperature
thermal density matrix as well as the zero-temperature ground state
considered so far. Note that the system is still thermally isolated
during the evolution, i.e., it is decoupled from a finite-temperature
heat bath at time $t=0$.

Let us briefly review the formalism. \cite{Polkovnikov} In terms
of phase-space variable $x$ and $p$ (which can be vectors for multi-dimensional
problems), the Weyl symbol $\Omega_{w}(x,p)$ for a quantum operator
$\hat{\Omega}$ is defined as 
\begin{equation}
\Omega_{w}({x},{p})=\int d{s}\left<{x}-\frac{{s}}{2}\right|\hat{\Omega}\left|{x}+\frac{{s}}{2}\right>\exp\left[\frac{i}{\hbar}{p}\cdot{s}\right].
\end{equation}
 The Wigner function $W(x,p)$ is, by definition, the Weyl symbol
for the density matrix $\hat{\rho}$, which satisfies the following
equation of motion: 
\begin{equation}
\partial_{t}W=-\frac{2}{\hbar}H_{w}\sin\left(\frac{\hbar}{2}\Lambda\right)W,\label{MB}
\end{equation}
 where $H_{w}$ is the Weyl symbol for the Hamiltonian, and $\Lambda\equiv\overleftarrow{\partial_{p}}\,\,\overrightarrow{\partial_{x}}-\overleftarrow{\partial_{x}}\,\,\overrightarrow{\partial_{p}}$.
The expectation value of an operator $\hat{O}$ can be written as
\begin{equation}
\langle\hat{O}\rangle=\iint_{-\infty}^{+\infty}\frac{dxdp}{2\pi}\: W(x,p)\,\, O_{w}(x,p)\label{expect}
\end{equation}
 in terms of its Weyl symbol $O_{w}(x,p)$ and the Wigner function
$W(x,p)$.

Let us now consider the single-mode Hamiltonian $\hat{H_{q}}(K(t))=\left(\frac{K(t)}{4}\:\hat{p}^{2}+\frac{1}{K(t)}\: q^{2}\hat{x}^{2}\right)$
[see Eq.~\eqref{eq:H-momentum-space}], where we have set $u=1$
and suppressed the $\Re$ or $\Im$ superscripts. Here, $\hat{x}$
and $\hat{p}$ respectively represent the real (or imaginary) part
of operators $\Phi_{q}$ and $\Pi_{q}$, and we have used the hat
notation to distinguish quantum operators from phase-space variables.
For a system evolving from an initial thermal density matrix, 
\begin{equation}
\hat{\rho}_{0}=\exp\left(-\hat{H_{q}}(K_{0})/k_{B}T_{0}\right)/{\rm tr}\left[\exp\left(-\hat{H_{q}}(K_{0})/k_{B}T_{0}\right)\right],
\end{equation}
 with the Hamiltonian above (for arbitrary $K(t)$), the Wigner function
retains the following form: 
\begin{equation}
W(x,p)={\cal N}\exp\left[-\frac{A}{2}x^{2}-\frac{B}{2}p^{2}+Cxp\right],\label{wigner}
\end{equation}
 where $A$, $B$, $C$, and ${\cal N}$ are potentially time-dependent functions
with the following initial conditions: 
\begin{equation}
\begin{split} & A(t=0)=\frac{4q}{K_{0}R},\quad B(t=0)=\frac{K_{0}}{qR},\quad C(t=0)=0,\\
 & {\cal N}(t=0)=\frac{2}{R},\quad R=\coth\left(\frac{q}{2k_{B}T_{0}}\right).
\end{split}
\label{IC}
\end{equation}
 Note that the normalization condition $\iint_{-\infty}^{+\infty}\frac{dxdp}{2\pi}W(x,p)=1$
sets 
\begin{equation}
{\cal N}=\sqrt{AB-C^{2}}\label{N}
\end{equation}
 at all times. For the quadratic Hamiltonian $\hat{H_{q}}(K(t))$,
the Weyl symbol is simply obtained by replacing the operators $\hat{x},\hat{p}$
with the phase-space variable $x,p$. We can then write Eq.~\eqref{MB}
as 
\[
\begin{split} & W(x,p)\left(\frac{\dot{{\cal N}}}{{\cal N}}-\frac{\dot{A}}{2}x^{2}-\frac{\dot{B}}{2}p^{2}+\dot{C}xp\right)\\
 & =W(x,p)\left(2\frac{q^{2}}{K(t)}\: x\:(Cx-Bp)-\frac{K(t)}{2}\: p\:(Cp-Ax)\right),
\end{split}
\]
 which leads to the following equations for the parameters of the
Wigner function: 
\begin{equation}
\begin{split} & \dot{{\cal N}}=0,\quad\dot{A}=-4\frac{q^{2}}{K(t)}C,\quad\dot{B}=K(t)C,\\
 & \dot{C}=2\left(\frac{K(t)}{4}A-\frac{q^{2}}{K(t)}B\right).
\end{split}
.
\end{equation}
 By using Eq.~\eqref{N}, and noting that ${\cal N}$ does not change
from its initial value, we can eliminate $A$ from the equations above:
\begin{equation}
\dot{B}=K(t)C,\quad\dot{C}=K(t)\frac{(4+C^{2}R^{2})}{2R^{2}B}-\frac{2q^{2}B}{K(t)}.\label{small-system}
\end{equation}
 We now expand Eq.~\eqref{small-system} in $\delta K(t)\ll K_{0}$,
and obtain two coupled Langevin equations: 
\begin{equation}
\begin{split}\dot{B} & =CK_{0}-CK_{0}^{2}\delta\alpha(t),\\
\dot{C} & =\frac{K0(4+C^{2}R^{2})}{2R^{2}B}-\frac{2q^{2}B}{K_{0}}-\left[2Bq^{2}+\frac{K_{0}^{2}(4+C^{2}R^{2})}{2BR^{2}}\right]\delta\alpha(t),
\end{split}
\label{wig_langevin}
\end{equation}
 where $\delta\alpha(t)=-\delta K(t)/K_{0}^{2}$. Using Eqs.~\eqref{expect}
and~\eqref{IC} , we can also express the absorbed energy in terms
$B$ and $C$ as 
\begin{equation}
\langle\hat{H}_{q}\rangle-\langle\hat{H}_{q}\rangle_{t=0}=\frac{K_{0}(4+C^{2}R^{2})}{16B}+\frac{q^{2}R^{2}B}{4K_{0}}-\frac{qR}{2}.\label{wig_e}
\end{equation}
 As expected, for $R=1$ ($T_{0}=0$), Eqs.~\eqref{wig_langevin}
and \eqref{wig_e} above are respectively equivalent to Eqs.~\eqref{nl_lang}
and \eqref{eq:average-FP} through a change of variables: $B=1/q\Re z_{q}$
and $C=-2\Im z_{q}/\Re z_{q}$. Applying the Fokker-Planck approach
of Appendix~\ref{sec:Fokker-Plank-approach-for} to Eqs.~\eqref{wig_langevin}
and \eqref{wig_e}, we obtain $\overline{\langle\epsilon_{q}\rangle}=\frac{qR}{2}\left(\exp\left[2k_{0}^{2}W^{2}q^{2}t\right]-1\right)$,
and find that the noise-averaged absorbed energy depends on the initial
temperature only through the prefactor $R=\coth\left(\frac{q}{2k_{B}T_{0}}\right)$.


\begin{thebibliography}{10}
\bibitem{note} We can simply think of $f(\psi,t)$ as the joint probability
distribution for all the independent real numbers required to label
the wave function $\psi$ modulo an overall $U(1)$ phase. Similarly,
$\int d\psi$ is shorthand notation for the integration over all such
numbers.
\bibitem{H0} We assume that at the moment of measurement the noise
is turned off and $H(g(t))=H(g_{0})$. The effect of noise is encoded
in the wave function of the system, which has evolved with the noisy
Hamiltonian over the time $t$.
\end{thebibliography}
\end{document}